\documentclass[iop]{emulateapj}

\usepackage{natbib}

\usepackage[backref,breaklinks,colorlinks,citecolor=blue,linkcolor=blue,urlcolor=blue]{hyperref}
\usepackage[all]{hypcap}

\newcommand{\figurewidth}{0.99}

\newcommand{\msun}{\,\mbox{M}_\odot}

\newcommand{\mpc}{\,\mbox{Mpc}}

\newcommand{\kpc}{\,\mbox{kpc}}

\newcommand{\pc}{\,\mbox{pc}}

\newcommand{\gyr}{\,\mbox{Gyr}}

\newcommand{\omegamatter}{\Omega_{\rm matter}}

\newcommand{\omegalambda}{\Omega_{\rm \Lambda}}

\newcommand{\mpeak}{M_{\rm peak}}
\newcommand{\mhalo}{M_{\rm halo}}

\newcommand{\mvir}{M_{\rm vir}}
\newcommand{\rvir}{R_{\rm vir}}

\newcommand{\mstar}{M_{\rm star}}
\newcommand{\Mgas}{M_{\rm gas}}

\newcommand{\mhi}{M_{\rm HI}}
\newcommand{\mhtwo}{M_{\rm H_2}}

\begin{document}

\journalinfo{The Astrophysical Journal Letters, accepted}

\title{Rapid Environmental Quenching of Satellite Dwarf Galaxies in the Local Group}
\shorttitle{Rapid Quenching of Satellite Dwarf Galaxies}
\shortauthors{Wetzel, Tollerud \& Weisz}

\author{Andrew R. Wetzel\altaffilmark{1, 2}}
\altaffiltext{1}{TAPIR, California Institute of Technology, Pasadena, CA, USA}
\altaffiltext{2}{Carnegie Observatories, Pasadena, CA, USA}

\author{Erik J. Tollerud\altaffilmark{3, 5}}
\altaffiltext{3}{Department of Astronomy, Yale University, New Haven, CT, USA}
 
\author{Daniel R. Weisz\altaffilmark{4, 5}}
\altaffiltext{3}{Department of Astronomy, University of Washington, Seattle, WA, USA}
\altaffiltext{5}{Hubble Fellow}

\begin{abstract}
In the Local Group, nearly all of the dwarf galaxies ($\mstar\lesssim10^9\msun$) that are satellites within $300\kpc$ (the virial radius) of the Milky Way (MW) and Andromeda (M31) have quiescent star formation and little-to-no cold gas.
This contrasts strongly with comparatively isolated dwarf galaxies, which are almost all actively star-forming and gas-rich.
This near dichotomy implies a \textit{rapid} transformation of satellite dwarf galaxies after falling into the halos of the MW or M31.
We combine the observed quiescent fractions for satellites of the MW and M31 with the infall times of satellites from the Exploring the Local Volume in Simulations (ELVIS) suite of cosmological zoom-in simulations to determine the typical timescales over which environmental processes within the MW/M31 halos remove gas and quench star formation in low-mass satellite galaxies.
The quenching timescales for satellites with $\mstar<10^8\msun$ are short, $\lesssim2\gyr$, and quenching is more rapid at lower $\mstar$.
These satellite quenching timescales can be $1-2\gyr$ longer if one includes the time that satellites were environmentally preprocessed by low-mass groups prior to MW/M31 infall.
We compare with quenching timescales for more massive satellites from previous works to synthesize the nature of satellite galaxy quenching across the observable range of $\mstar=10^{3-11}\msun$.
The satellite quenching timescale increases rapidly with satellite $\mstar$, peaking at $\approx9.5\gyr$ for $\mstar\sim10^9\msun$, and the timescale rapidly decreases at higher $\mstar$ to $<5\gyr$ at $\mstar>5\times10^9\msun$.
Overall, galaxies with $\mstar\sim10^9\msun$, similar to the Magellanic Clouds, exhibit the longest quenching timescales, regardless of environmental or internal mechanisms.
\end{abstract}


\section{Introduction}

Galaxies in denser environments are more likely to have suppressed (quiescent) star formation and little-to-no cold gas than galaxies of similar stellar mass, $\mstar$, in less dense environments.
The observed environmental effects within the Local Group (LG), on the satellite galaxies within the halos of the Milky Way (MW) and Andromeda (M31), are particularly strong \citep[e.g.,][]{Einasto1974, GrcevichPutman2009, McConnachie2012, Phillips2014, SlaterBell2014}, even compared to the already strong effects on (more massive) satellites within massive groups/clusters \citep[e.g.,][]{Wetzel2012}.
Specifically, dwarf galaxies around the MW/M31 show a strikingly sharp and nearly complete transition in their properties within $\approx 300 \kpc$ (approximately the virial radius, $\rvir$, of the MW or M31), from irregular to spheroidal morphologies, from significant to little-to-no cold atomic gas, and from star-forming to quiescent.
This trend has just a few exceptions: 4 gas-rich, star-forming galaxies persist within the halos of the MW (the LMC and SMC) and M31 (LGS 3 and IC 10), and 4 - 5 quiescent, gas-poor galaxies reside well beyond $\rvir$ of either the MW or M31: Cetus \citep{Lewis2007}, Tucana \citep{Fraternali2009}, KKR 25 \citep{Makarov2012}, KKs 3 \citep{Karachentsev2015}, and possibly Andromeda XVIII, though Cetus and Tucana may have orbited within the MW halo \citep{Teyssier2012}.
This efficient satellite quenching is particularly striking because, other than KKR 25 and KKs 3, at $\mstar<10^9\msun$ all known galaxies that are sufficiently isolated ($>1500\kpc$ from a more massive galaxy) are star-forming \citep{Geha2012, Phillips2014}.
Thus, the MW and M31 halos show the strongest environmental influence over their satellites of any known systems, making the LG a compelling laboratory for studying environmental processes on galaxies.

Several such processes within a host halo can regulate the gas content, star formation, morphology, and eventual disruption of satellites, including gravitational tidal forces \citep[e.g.,][]{Dekel2003}, galaxy--galaxy tidal interactions \citep[e.g.,][]{FaroukiShapiro1981}, galaxy--galaxy mergers \citep[e.g.,][]{Deason2014a}, and ram-pressure stripping of extended gas \citep[e.g.,][]{McCarthy2008} or inter-stellar medium \citep[e.g.,][]{GunnGott1972, Tonnesen2009}.
The key astrophysical challenge is understanding the relative importance of these, including which (if any) dominate, and how they vary across both satellite and host masses.

One strong constraint comes from determining the timescale over which environmental quenching occurs, as previous works explored at higher masses \citep[e.g.,][]{Balogh2000, DeLucia2012, Wetzel2013, Hirschmann2014, Wheeler2014}.
For the satellite dwarf galaxies in the LG, recent works showed that their environmental quenching \textit{efficiency} is higher than for higher-mass satellites \citep{Phillips2014, SlaterBell2014}.
In this letter, we combine the observed quiescent fractions for satellites in the MW/M31 halos with their typical infall times from cosmological simulations to infer the timescales over which environmental processes remove their gas and quench star formation.
Motivated by \citet{Wetzel2015a}, we also consider the possible impact of group preprocessing on satellites before they fell into the MW/M31 halos.
We also compare with previous works on more massive satellites, to synthesize satellite quenching across the observable range of $\mstar=10^{3-11}\msun$.

\section{Methods}

\subsection{Observations}

\begin{figure}
\centering
\includegraphics[width = \figurewidth \columnwidth]{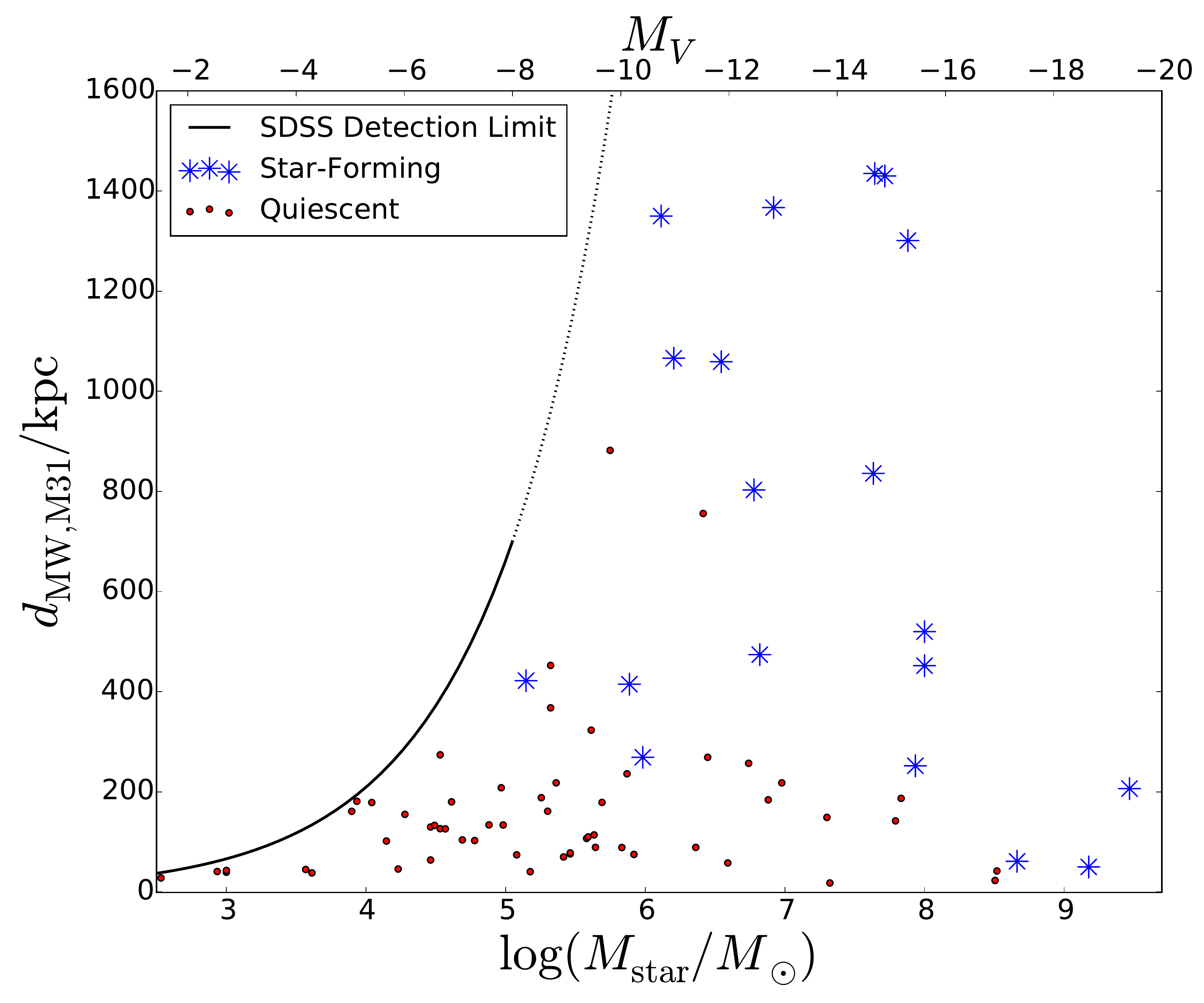}
\caption{
For all known dwarf galaxies in the Local Group out to $1.6\mpc$, the distance from their nearest host (MW or M31) versus stellar mass, $\mstar$, or absolute magnitude, $M_V$.
Points show individual galaxies: actively star-forming (blue stars) and quiescent ($\Mgas/\mstar<0.1$, red circles).
Black curve shows the detection limit for dwarf spheroidal-like galaxies with stars resolved in SDSS, plus extrapolation beyond $\sim700\kpc$.
}
\label{fig:local_group}
\end{figure}

To examine the observed properties of dwarf galaxies in the LG, we use the compilation from \citet{McConnachie2012}, which includes all galaxies known at that time within $3\mpc$ of the Sun.
We also include the more recent observations of cold atomic gas mass from \citet{Spekkens2014}.
We define ``satellite'' galaxies as those within $300\kpc$ of either the MW or M31, motivated by the observed sharp transition in star formation, gas mass, and morphology within this distance.

Observed dwarf galaxies show a tight correlation between morphology, star formation, and cold gas mass: all spheroidals have little-to-no detectable cold gas \citep[e.g.,][]{Spekkens2014} or star formation \citep[e.g.,][]{Weisz2014a}, and almost all irregulars have significant cold gas and ongoing star formation.
Thus, we define ``quiescent'' galaxies as having $\Mgas/\mstar<0.1$ or, if they have no cold gas constraints, having colors/morphologies that resemble spheroidals.
By this definition, the only star-forming, gas-rich satellites are: LMC ($\mstar=1.5\times10^9\msun$, $\Mgas/\mstar\approx0.3$) and SMC ($\mstar=4.6\times10^8\msun$, $\Mgas/\mstar\sim1$) around the MW, LGS 3 ($\mstar=9.6\times10^5\msun$, $\Mgas/\mstar\approx0.4$) and IC 10 ($\mstar=9\times10^7$, $\Mgas/\mstar\approx0.6$) around M31.

For each dwarf galaxy out to $1.6\mpc$, Figure~\ref{fig:local_group} shows its distance from nearest host (MW or M31) versus $\mstar$.
Almost all quiescent galaxies are within $\approx300\kpc$ of their host.
The black curve shows the detection limit (and extrapolation) for dwarf spheroidal-like galaxies in SDSS \citep{Tollerud2008}, which highlights completeness at different $\mstar$.

\subsection{Simulations}

To measure the infall times of satellites, we use Exploring the Local Volume in Simulations (ELVIS), a suite of cosmological zoom-in $N$-body simulations intended to model the LG \citep{GarrisonKimmel2014} in $\Lambda$CDM cosmology: $\sigma_8=0.801$, $\omegamatter=0.266$, $\omegalambda=0.734$, $n_s=0.963$ and $h=0.71$.
Within the zoom-in regions, the particle mass is $1.9\times10^5\msun$ and the Plummer-equivalent force softening is $140\pc$ physical.

ELVIS contains 48 dark-matter halos of masses similar to the MW or M31 ($\mvir=1.0-2.8\times10^{12}\msun$), with a median $\rvir\approx300\kpc$.
Half of the halos are in a pair that resemble the masses, distance, and relative velocity of the MW--M31 pair, while the other half are single isolated halos.
Given the lack of systematic differences in satellite infall times for the paired versus isolated halos \citep{Wetzel2015a}, we use all 48 to improve statistics.

ELVIS identifies dark-matter (sub)halos using the six-dimensional halo finder \textsc{rockstar} \citep{Behroozi2013a}.
For each halo, we assign a virial mass, $\mvir$, and radius, $\rvir$, according to \citet{BryanNorman1998}.
A ``subhalo'' is a halo whose center is inside $\rvir$ of a more massive host halo, and a subhalo experiences ``first infall'' and becomes a ``satellite'' when it \textit{first} passes within $\rvir$.
For each subhalo, we compute the peak mass, $\mpeak$, that it ever reached, and we assign $\mstar$ to subhalos based on $\mpeak$ using the relation from abundance matching in \citet{GarrisonKimmel2014}, which reproduces the observed mass function in the LG if one accounts for observational incompleteness \citep{Tollerud2008, Hargis2014}.

For more on ELVIS and its satellites' infall times, see \citet{GarrisonKimmel2014} and \citet{Wetzel2015a}.

\section{Results}

\subsection{Observed Quiescent Fractions for Satellites}

\begin{figure}
\centering
\includegraphics[width = \figurewidth \columnwidth]{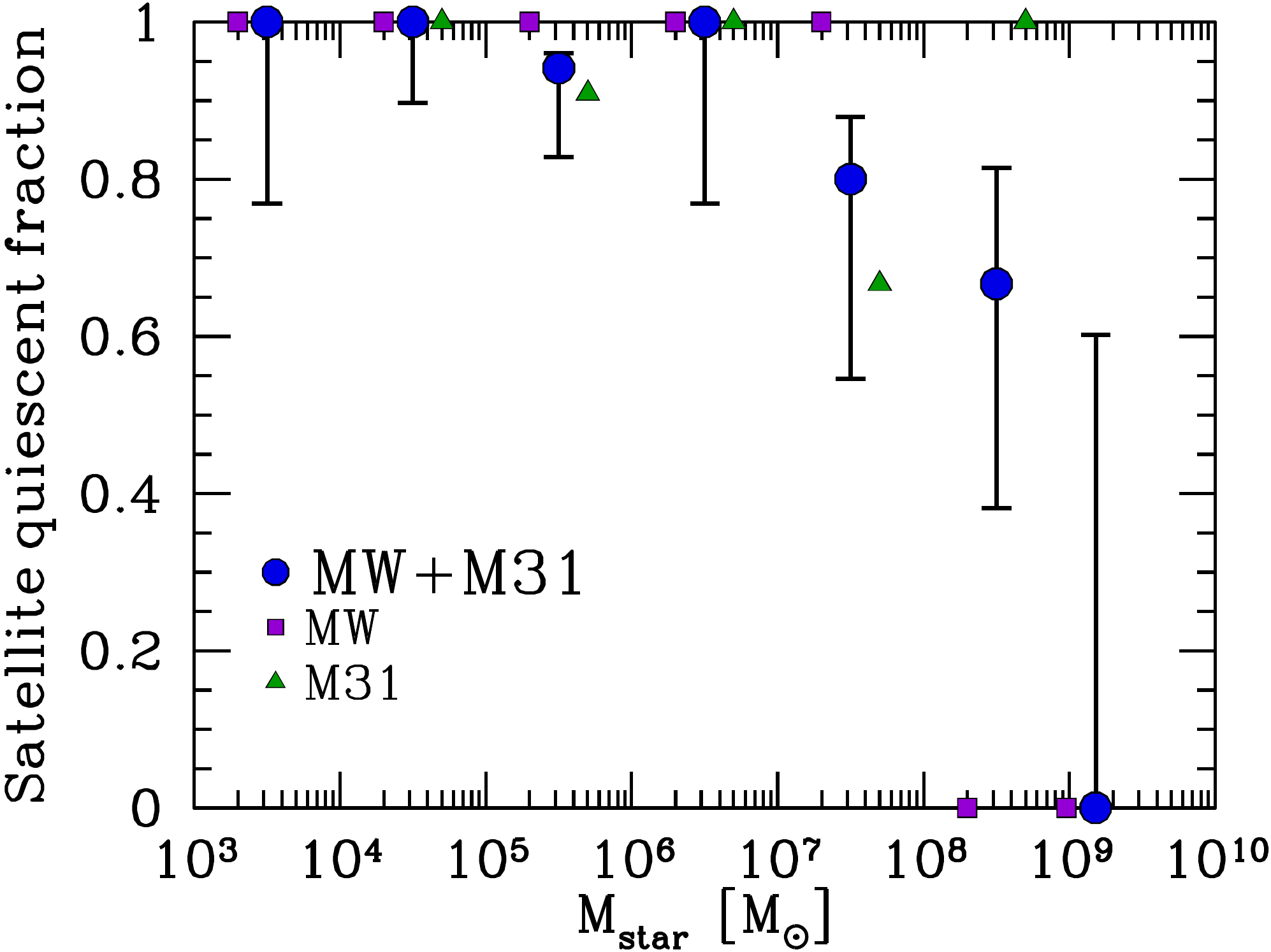}
\caption{
For all satellites galaxies with $\mstar\lesssim10^9\msun$ within $300\kpc$ of the Milky Way (MW) or Andromeda (M31), the fraction that are quiescent ($\Mgas/\mstar<0.1$) versus stellar mass, $\mstar$.
Blue circles show all satellites, violet squares (green triangles) show those of just the MW (M31).
Of these 56 satellites, only 4 are star-forming/gas-rich: LMC ($\mstar=1.5\times10^9\msun$) and SMC ($\mstar=4.6\times10^8\msun$) around the MW, LGS 3 ($\mstar=9.6\times10^5\msun$) and IC 10 ($\mstar=9\times10^7\msun$) around M31.
At $\mstar<8\times10^7\msun$, 50 of 51 satellites are quiescent, and at $\mstar<9\times10^5\msun$ \textit{all} are quiescent.
Error bars show 68\% uncertainty from observed counts.
}
\label{fig:quiescent_fraction}
\end{figure}

Figure~\ref{fig:quiescent_fraction} shows, for all satellite galaxies at $\mstar\lesssim10^9\msun$ within $300\kpc$ of the MW or M31, the fraction that are quiescent, in 1-dex bins of $\mstar$ \citep[see also][]{Phillips2014, SlaterBell2014}.
We do not correct for observational completeness versus $\mstar$ (Figure~\ref{fig:local_group}), because we measure the relative fraction in each bin, which is likely unbiased.
We show fractions for all satellites (blue circles) and separately for those in the MW (violet squares) and M31 (green triangles) halos.
Error bars show 68\% uncertainty for the binomial counts using a beta distribution.
Of the 56 satellites, only 4 (7\%) are star-forming/gas-rich: LMC and SMC of the MW, LGS 3 and IC 10 of M31.
Moreover, at $\mstar<8\times10^7\msun$, only 1 (LGS 3) of the 51 satellites is star-forming, and at $\mstar<9\times10^5\msun$ \textit{all} 40 satellites are quiescent.

These near-unity quiescent fractions for satellites of the MW/M31 contrast strongly with the nearly \textit{zero} quiescent fraction for isolated (non-satellite) galaxies at $\mstar<10^9\msun$ \citep{Geha2012, Phillips2014}.
The only clear exceptions are the quiescent galaxies KKR 25 ($\mstar=1.4\times10^6\msun$) and KKs 3 ($\mstar=2.3\times10^7\msun$) at $\approx2\mpc$ from the MW/M31.
(Though, as Figure~\ref{fig:local_group} shows, the completeness distances at low $\mstar$ leave open the possibility for more isolated quiescent dwarf galaxies.)

\subsection{Inferred Quenching Timescales for Satellites}

\renewcommand{\figurewidth}{0.48}
\begin{figure*}
\centering
\includegraphics[width = \figurewidth \textwidth]{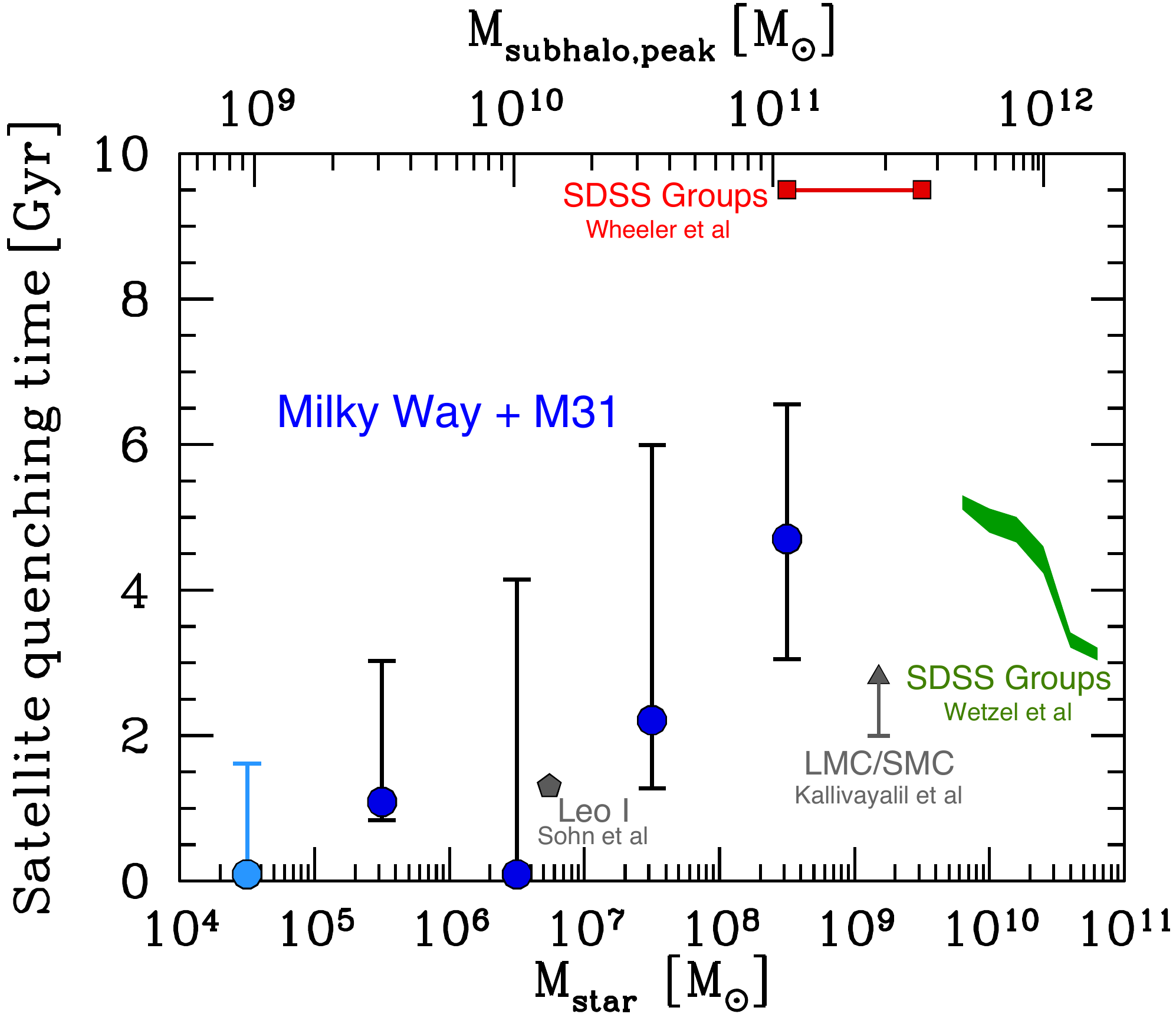}
\hspace{3 mm}
\includegraphics[width = \figurewidth \textwidth]{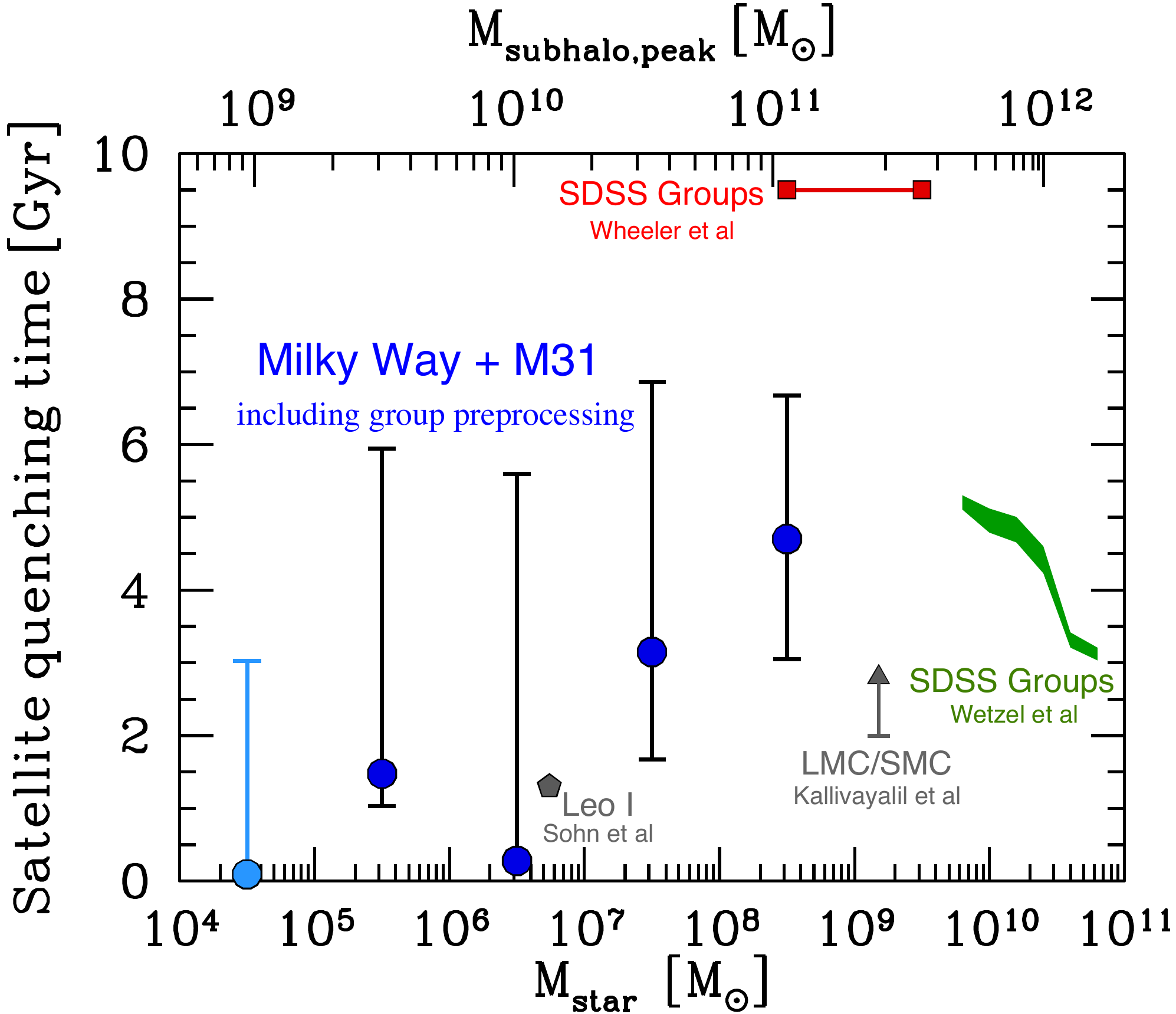}
\caption{
Satellite quenching timescales for galaxies across the observable range of stellar mass, $\mstar$ (top axis shows subhalo $\mpeak$ from abundance matching).
Blue circles show satellites of the MW and M31, obtained by matching the observed quiescent fractions in Figure~\ref{fig:quiescent_fraction} to rank-ordered infall times of satellites from the ELVIS simulations \citep{Wetzel2015a} in 1-dex bins of $\mstar$.
At $\mstar=10^{4-5}\msun$ (light blue), reionization may have quenched some satellites prior to infall.
Error bars come from the 68\% uncertainty in observed quiescent fractions in Figure~\ref{fig:quiescent_fraction}.
Left panel uses time since first infall into the current MW/M31-like halo, while right panel uses time since first infall into \textit{any} host halo, thereby including possible effects of group preprocessing.
Gray triangle shows lower limit for the LMC/SMC system from its measured orbit \citep{Kallivayalil2013}, and gray pentagon shows the quenching timescale for Leo I from its measured orbit and star-formation history \citep{Sohn2013}.
Red squares show times inferred for satellites with $\mstar=10^{8.5}$, $10^{9.5}\msun$ around hosts with $\mstar>2.5\times10^{10}\msun$ in SDSS \citep{Wheeler2014}, and green curve shows the same for more massive satellites in groups of $\mvir=10^{12-13}\msun$ in SDSS \citep{Wetzel2013}.
The satellites in the MW/M31 halos quenched more rapidly after infall than more massive satellites (around other hosts).
Overall, the quenching timescale increases with $\mstar$, is longest at $\mstar\sim10^9\msun$ (near the masses of the Magellanic Clouds), then decreases with further increasing $\mstar$.
}
\label{fig:quench_times}
\end{figure*}

We now translate the quiescent fractions in Figure~\ref{fig:quiescent_fraction} into the typical timescales over which environmental processes quenched satellites of the MW/M31 after they fell into a host halo, following the methodology of \citet{Wetzel2013}.

First, motivated by the dearth of \textit{isolated} galaxies with $\mstar<10^9\msun$ that are quiescent at $z\approx0$, we assume that all satellites with $\mstar(z=0)<10^9\msun$ were star-forming prior to first infall.
However, we do not model $\mstar(z=0)<10^4\msun$, because cosmic reionization likely quenched most/all such galaxies at high redshift \citep[e.g.,][]{Weisz2014a, Brown2014}.
At $\mstar(z=0)=10^{4-5}\msun$, satellites' star-formation histories show a mix of complete quenching by $z\gtrsim3$ (e.g., Bootes I, Leo IV) and signs of star formation at $z\lesssim1$ (e.g., And XI, And XII, And XVI) \citep{Weisz2014a, Weisz2014c, Brown2014}, so quenching at these masses may arise from a mix of reionization and the host-halo environment.
That said, the 100\% quiescent fraction for satellites at this $\mstar$ means that if both processes are responsible, both are highly efficient.
Furthermore, if the satellites that were quenched by reionization versus the host-halo environment have similar infall-time distributions, our modeling approach remains valid.
Thus, we include this $\mstar$ but label it distinctly to emphasize caution in interpretation.
 
Within each 1-dex bin of $\mstar$, we use ELVIS to compute the distribution of infall times for satellites at $z=0$.
Infall into the MW/M31 halo (or any host halo) typically occurred $5-8\gyr$ (or $7-10\gyr$) ago, and our most massive satellites typically fell in $2-3\gyr$ more recently than our least massive \citep[see Figures 1 and 2 in][]{Wetzel2015a}.
Assuming that environmental quenching correlates with time since infall, we designate those that fell in earliest as having quenched, and we adjust the time-since-infall threshold for quenching until we match the observed quiescent fraction in each $\mstar$ bin.

Several works have shown that this model successfully describes the dependence of satellite quiescent fractions on host-centric distance \citep[e.g.,][]{Wetzel2013, Wetzel2014, Wheeler2014} because infall time correlates with host-centric distance \citep[e.g.,][]{Wetzel2015a}.
However, this correlation means that we must account for the distances of the observed satellites in computing their infall times.
Thus, in ELVIS we only select satellites out to the maximum host-centric distance that they are observed in each $\mstar$ bin.
This matters most at the highest $\mstar$, where all observed satellites (M32, NGC 205, LMC/SMC) reside $\lesssim60\kpc$ from the MW or M31.

Figure~\ref{fig:quench_times} shows the resultant environmental quenching timescales (the time duration from first infall to being fully quiescent/gas-poor) for satellites versus their $\mstar$ (or subhalo $\mpeak$).
Blue circles show satellites in the MW and M31, and we shade the lowest $\mstar$ to highlight caution in interpretation because of reionization.
We derive error bars from the 68\% uncertainty in the observed quiescent fractions in Figure~\ref{fig:quiescent_fraction}.

As explored in \citet{Wetzel2015a}, many satellites first fell into a another host halo (group), typically with $\mvir\sim10^{11}\msun$, before falling into the MW/M31 halos.
Such groups may correspond to, for example, the LMC, as the newly discovered dwarf galaxies near the LMC \citep{Koposov2015, Bechtol2015} suggest, although such groups disperse in phase space $\sim5\gyr$ after MW/M31 infall \citep{Deason2015}, and typically half of such preprocessing hosts do not survive to $z=0$ \citep{Wetzel2015a}, so preprocessed satellites are not always easily distinguishable.
Because the importance of this preprocessing in low-mass groups remains unclear, we present quenching timescales both neglecting (left panel) and including (right panel) group preprocessing.
The latter results in longer quenching timescales, though it primarily shifts the upper tail of the distribution and not the median.

Both panels show shorter median quenching timescales for less massive satellites: $\sim5\gyr$ at $\mstar=10^{8-9}\msun$, $2-3\gyr$ at $\mstar=10^{7-8}\msun$, and $<1.5\gyr$ at $\mstar<10^7\msun$, depending on group preprocessing.
Moreover, the median timescale for two of the lowest $\mstar$ bins is $0\gyr$, because 100\% of those satellites are quiescent, which implies extremely rapid quenching after infall.

Figure~\ref{fig:quench_times} also shows infall/quenching timescales that are more directly measured for satellites of the MW.
The 3-D orbital velocity measured for the LMC/SMC strongly suggests that they are experiencing first infall and crossed inside $\rvir$ of the MW $\approx2\gyr$ ago \citep{Kallivayalil2013}.
Given that the LMC and SMC remain star-forming, this places a lower limit to their quenching timescale (gray triangle), consistent with our statistical timescales.
Similarly, measurements of the 3-D orbital velocity and star-formation history for Leo I indicate that it fell into the MW halo $\approx2.3\gyr$ ago and quenched $\approx1\gyr$ ago (near its $\approx90\kpc$ pericentric passage), implying a quenching timescale of $\approx1.3\gyr$ \citep[][gray pentagon]{Sohn2013}, again consistent with our results.

The mass trend in Figure~\ref{fig:quench_times} is broadly consistent the star-formation-history-based results of \citet{Weisz2015} that more massive dwarf galaxies in the LG quenched more recently.
Also, the overall timescale is broadly consistent with \citet{SlaterBell2014}, who inferred a typical quenching time since first \textit{pericenter} of $1-2\gyr$, which implies a quenching time since \textit{infall} of $\sim3\gyr$, though they did not examine mass dependence.

We also compare these timescales with previous results for more massive satellites of other hosts.
Figure~\ref{fig:quench_times} (green curve) shows the quenching timescales for satellites in groups with $\mvir=10^{12-13}\msun$ from \citet{Wetzel2013}, who used identical methodology based on galaxies in SDSS \citep{Tinker2011, Wetzel2012}.
(\citet{Hirschmann2014} found similar timescales versus $\mstar$.)
Red squares show timescales from \citet{Wheeler2014}, who also used an SDSS galaxy catalog \citep{Geha2012} and similar methodology for satellites with $\mstar\approx10^{8.5 - 9.5}\msun$ around hosts with $\mstar>2.5\times10^{10}\msun$, or $\mvir\approx10^{12.5-14}\msun$, much more massive than the MW/M31.
Both works measured satellite infall times, including group preprocessing, from cosmological simulations.
The timescale changes rapidly between these works, from $\approx5.2\gyr$ at $\mstar\approx10^{9.8}\msun$ to $\approx9.5\gyr$ at $\approx10^{9.5}\msun$.
Both analyses used similar galaxy catalogs and methodologies, though \citet{Wetzel2013} used a group catalog to narrow the masses of the hosts, which are more similar to the MW/M31, while the hosts in \citet{Wheeler2014} are more massive, on average.
Thus, part of this change in timescale could arise if more massive hosts quench satellites with $\mstar\sim10^9\msun$ \textit{less} rapidly.
Absent that, these results imply that the satellite quenching timescale rises rapidly near $\mstar\sim10^9\msun$.
Furthermore, the timescale from \citet{Wheeler2014} implies some tension with our $\approx5\gyr$ at $\mstar\approx10^{8.5}\msun$, as driven by the higher quiescent fraction in the MW/M31 at this $\mstar$, specifically, NGC 205 and M32, two quiescent satellites of M31.
This tension could be explained if NGC 205 and M32 both fell into the M31 halo unusually early ($>9.5\gyr$ ago), and/or (again) if M31 quenches its satellites more rapidly than the higher-mass ($\mvir=10^{12.5-14}\msun$) host halos in \citet{Wheeler2014}.

Altogether, Figure~\ref{fig:quench_times} indicates a complex dependence of the satellite quenching timescale on $\mstar$.
The typical timescale for satellites of the MW/M31 increases with $\mstar$, from $\lesssim1\gyr$ at $\mstar<10^7\msun$ to $\sim5\gyr$ at $\mstar\approx10^{8.5}\msun$.
\citet{Wheeler2014} indicates that this mass dependence continues, though with a rapid increase ($\sim2\times$) to $\approx9.5\gyr$, and no change from $\mstar\approx10^{8.5}$ to $10^{9.5}\msun$.
Finally, \citet{Wetzel2013} shows that the timescale \textit{decreases} near $5\times10^9\msun$ and continues to decline with increasing $\mstar$.
Overall, the typical satellite quenching timescale is shortest at lowest $\mstar$, short at the highest $\mstar$, and longest at $\mstar\sim10^9\msun$, comparable to the Magellanic Clouds.

\section{Discussion}

We conclude by discussing the dependence of satellite quenching timescales on $\mstar$ from Figure~\ref{fig:quench_times} in the context of the underlying physics.

At $\mstar\gtrsim10^9\msun$, the long timescales suggests that satellite quenching is caused by gas depletion in the absence of cosmic accretion, via the stripping of extended gas around a satellite after infall (``strangulation'').
This scenario can explain shorter timescales at increasing $\mstar$, because higher-$\mstar$ star-forming galaxies have lower $\Mgas/\mstar$ \citep[in cold atomic and molecular gas, e.g.,][]{Huang2012, Boselli2014} and thus shorter gas depletion timescales in the absence of accretion.
For example, \citet{Bradford2015} found that isolated galaxies follow $\Mgas/\mstar\propto\mstar^{-0.55}$ at $\mstar>10^{8.6}\msun$.
Furthermore, star-forming galaxies at $\mstar\sim10^9\msun$ have $\Mgas=\mhi+\mhtwo\approx\mstar$, with gas depletion timescales comparable to a Hubble time.
Thus, satellite quenching timescales at $\mstar\gtrsim10^9\msun$ do not necessarily \textit{require} strong environmental processes beyond truncated gas accretion \citep[see also][]{Wetzel2013, Wheeler2014, McGee2014}.
Furthermore, at $\mstar>10^9\msun$, internal feedback from stars and/or black holes also may quench satellites after infall, which could help explain the shortening of the timescale with increasing $\mstar$.

However, strangulation cannot explain the rollover in quenching times at $\mstar\lesssim10^9\msun$, because the star-forming dwarf galaxies in the LG also have $\Mgas\gtrsim\mstar$ \citep{GrcevichPutman2009}, enough to fuel star formation for a Hubble time.
Thus, the rapid decline of the timescale at lower $\mstar$ \textit{requires} an additional process(es) to remove gas from satellites after infall.
This likely arises from increased efficiency of ram-pressure stripping in removing cold gas from such low-mass galaxies, which have shallower potential wells.
Moreover, the same internal stellar feedback that regulates the low star-formation efficiency in dwarf galaxies likely heats/drives significant cold gas to large radii \citep[e.g.,][]{Muratov2015}, which would assist such environmental stripping.
Thus, the rapid quenching timescales for dwarf galaxies may arise from the nonlinear interplay of both internal feedback and external stripping \citep[e.g.,][]{NicholsBlandHawthorn2011, BaheMcCarthy2015}.

Overall, satellites with $\mstar\sim10^9\msun$ (similar to Magellanic Clouds) represent the transition between quenching via gas consumption and via gas stripping, and \textit{no} quenching mechanism, either internal or external, appears to operate efficiently near this mass \citep[see also][]{Geha2012, Weisz2015}.

Finally, the above scenario may explain the curious similarity between the mass dependence of the quenching timescale in Figure~\ref{fig:quench_times} and the underlying galaxy-halo $\mstar/\mhalo$ ratio, which also is small at both high and low $\mstar$ and peaks at $\mstar\sim10^{10}\msun$ \citep[e.g.,][]{Behroozi2013c}.
In particular, at high $\mstar$, the same physical process(es) that lowers $\mstar/\mhalo$ also lowers a galaxy's cold gas mass, which in turn causes more massive satellites to quench more rapidly, absent accretion.
At low $\mstar$, the same shallower potential well that allows stellar feedback to lower $\mstar/\mhalo$ also allows external stripping to occur more easily and thus quenching to occur more rapidly.

During preparation, we learned of \citet{Fillingham2015}, who also used ELVIS to constrain the quenching timescales of satellites of the MW/M31 and reached similar conclusions.
\\

We thank the Aspen Center for Physics and the Kavli Institute for Theoretical Physics, both supported by the National Science Foundation, for stimulating environments.
A.~R.~W. gratefully acknowledges support from the Moore Center for Theoretical Cosmology and Physics at Caltech.
Support for E.~J.~T. and D.~R.~W. is provided by NASA through Hubble Fellowship grants HST-HF-51316.01 and HST-HF-51331.01, respectively.


\end{document}